\documentclass[12pt]{article}

\usepackage{amsmath}
\usepackage{amssymb}
\usepackage{amsfonts}
\usepackage{amscd}
\usepackage{amsbsy}
\usepackage{amsthm}
\usepackage{latexsym}%%
\usepackage{graphicx}
\usepackage{slashed}
\usepackage{color}

\DeclareMathOperator{\cx}{\square}
\renewcommand{\baselinestretch}{1.2}
\def\beq{\begin{eqnarray}}
\def\eeq{\end{eqnarray}}
\newcommand{\nn}{\nonumber}

\def\ln{\,\mbox{ln}\,}

\def\Tr{\,\mbox{Tr}\,}

\def\al{\alpha}
\def\be{\beta}

\def\ga{\gamma}
\def\de{\delta}
\def\vp{\varepsilon}

\def\la{\lambda}
\def\na{\nabla}

\def\si{\sigma}

\def\ph{\varphi}

\def\Ga{\Gamma}
\def\De{\Delta}

\def\Si{\Sigma}

\usepackage[symbol]{footmisc} % fix the problem with numeration of footnotes
\usepackage{float} %%%   Force figure placement in text
\usepackage{cite}  %%%   Mulitple reference citation
\usepackage{titlesec}

\usepackage{ulem} %%risca o texto selecionando ao meio por linha

\titleformat*{\section}{\large\bfseries}
\titleformat*{\subsection}{\normalsize\bfseries}
\begin{document}

%%%%%%%%%%%%%%%%%%%%%%%%%%%%%%
\begin{center}

{\large\bf
Local conformal symmetry and anomalies
\\
with antisymmetric tensor field}
\vskip 5mm

\textbf{Ilya L. Shapiro}%% ~$^{a}$
\footnote{
%%
%%    \quad
E-mail address: \ ilyashapiro2003@ufjf.br}
%% \ , \ \ \ On leave from Tomsk State Pedagogical University}

%%%%%%%%%%%%%%%%%%%%%%%%%%%%%
\vskip 5mm

{\sl Departamento de F\'{\i}sica, ICE,
Universidade Federal de Juiz de Fora
\\
Campus Universit\'{a}rio - Juiz de Fora, 36036-900, MG, Brazil}
\end{center}
%%%%%%%%%%%%%%%%%%%%%%%%%%%%%%

\vskip 6mm

\centerline{\textbf{Abstract}}
\vskip 1mm

\begin{quotation}
\noindent
We consider the trace anomaly, which results from the integration of
the massless conformal fermion field with the background of metric
and antisymmetric tensor fields. The non-local terms in the
anomaly-induced effective action do not depend on the scheme of
quantum calculations. On the other hand, total derivative terms in
the anomaly and the corresponding local part of the induced action
manifest scheme dependence and multiplicative anomaly.
\vskip 4mm

\vskip 3mm

\noindent
\textit{MSC:} \ %% Conformal anomaly;
81T10,  %%  	Model quantum field theories
81T15,  %%  Perturbative methods of renormalization applied to problems in quantum field theory
81T20   %%  	Quantum field theory on curved space or space-time backgrounds
81T50   %% 	Anomalies in quantum field theory

\end{quotation}
%%%%%%%%%%%%%%%%%%%%%%%%%%%%

%%%%%%%%%%%%%%%%%%%%%%%%%%%%
%%%%%%%%%%%%%%%%%%%%%%%%%%%%
%%%%%%%%%%%%%%%%%%%%%%%%%%%%
\section{Introduction}
\label{SecIntro}

Conformal anomaly in four spacetime dimensions
is a significant element of the quantum field
theory in curved space \cite{duff77,duff94}. One of the reasons is
that anomaly offers the simplest possible shortcut to derive one-loop
corrections to the classical action. In particular, such an important
application as Hawking radiation of black holes \cite{Hawking75}
can be derived using anomaly \cite{Christensen1977} and one can
even go further and classify the vacuum states in the vicinity of
black holes, by using a natural indefiniteness
in the anomaly-induced action \cite{balsan}. The corresponding
ambiguities emerge because such an
action includes Green functions of the artificial fourth-derivative
Paneitz operator \cite{Paneitz} (constructed earlier as part of the
conformal supergravity program \cite{FrTs-superconf,Fradkin1985}).
At least in part, these ambiguities are equivalent of adding an extra
nonlocal conformal invariant term
to the classical action. There is another ambiguity in the anomaly
and in the induced action, related to the freedom of adding local
nonconformal term $R^2$ to the classical action, which modifies
the total derivative term $\cx R$ in the anomaly \cite{anomaly-2004}
(see also previous discussion, e.g., in \cite{birdav}, \cite{duff94}
and further references therein).

The ambiguity related to $\cx R$ is not critical for the consistency
of the semiclassical theory because it concerns the vacuum part of
effective action and does not affect the renormalizability. However,
there is a similar ambiguity in the interacting field theories, such
as the models with metric-scalar background. In this case, the
ambiguity may affect the structure of anomaly in higher loop
approximations. The known examples include metric-scalars
\cite{anomaly-2006} and metric-torsion \cite{AtA} cases. In the
present work, we present one more example of an ambiguity in the
total derivative terms in anomaly, which is also related to the
multiplicative anomaly (MA). Historically, this anomaly was first
reported as a result of comparison of
$\Tr \ln A + \Tr \ln B$ vs $\Tr \ln (AB)$  using zeta-regularization
on the de Sitter space
\cite{Kontsevich1994,Elizalde1997,Elizalde1998}.
Soon it was realized that this framework is insufficient to observe
the difference between the two expressions because such a difference
is hidden behind the $\mu$-dependencies which emerge when the
divergences are subtracted \cite{Dowker1998,Evans1998}. The effect
of the $\mu$-dependencies persists even in the framework of
zeta-regularization, regardless in this scheme the divergences are
hidden \cite{Elizalde1995}. All this means that the MA can be
observed, in the first place, in the nonlocal part of effective
action, since the divergences can always be removed by local
counterterms and therefore the nonlocal terms are not directly
affected by the $\mu$-dependence. The first work where this type
of MA was reported \cite{QED-form} had a qualitative explanation
that the MA is an unavoidable consequence of the universality of the
general form of finite coefficients in the Schwinger-DeWitt expansion.
Each trace of the \ $\Tr \hat{a}_k(x,x')$ possesses universality in the
dimension $D=2k$ where it corresponds to the logarithmic UV
divergence in the proper-time integral. When we sum these
coefficients in $4D$, the universality is lost. Thus, MA does not
occur in the logarithmic divergences and, consequently, is expected
to hold in the finite part of the effective action of massive quantum
fields. The known examples of MA of this type belong to the fermion
determinants doubling \cite{QED-form,Peixoto2012} and to the more
sophisticated realization in the massive vector model \cite{BarKal}.

In the recent paper, \cite{AtA}, it was suggested another type of MA
that takes place only for massless conformal fields. In these theories,
the anomaly-induced action includes local non-conformal terms such
as $R^2$ in the purely metric background case. Since the one-loop
divergences are conformal \cite{tmf}, these terms do not suffer from
the $\mu$-dependencies and may produce a new kind of MA. The
example constructed in \cite{AtA} concerns the fermion on the
background of an axial vector field. It is worth noting that the
effect of local MA does not exist on a purely metric or metric-scalar
background. In this sense, the founding of the local MA in \cite{AtA}
is not trivial. The present communication reports on the second
example, with the metric and antisymmetric tensor field background.
%%%%%%%%%%%%%%%%%%%%%%%%

The study of antisymmetric fields attracted significant attention,
starting from the classical works \cite{OgiPolu67} and
\cite{KalbRamon}. These studies left an important imprint on
string theory and related areas (see, e.g.,
\cite{Pasti1995,Quevedo1996,Siegel1999,Buch2008}).
The gauge invariant theory of the antisymmetric field (Kalb and Ramon
model) describes the propagation of the irreducible antisymmetric
tensor representation of the Lorentz group. This model is free of
ghosts and unitary at the quantum level.
On the other hand, the coupling to fermions results in the
non-renormalizable quantum theory.

A qualitatively different version of the antisymmetric tensor field
theory was introduced by Avdeev and
Chizhov \cite{Avdeev1993}, but its basic form was known much
earlier from mathematical investigations of conformal operators
\cite{Branson} (see also \cite{Barbashov1983,Erdmenger1997}) and
from the conformal supergravity \cite{FrTs-superconf,Fradkin1985}.
In this case, there is no gauge symmetry. Thus, the conformal version
has more degrees of freedom, including unphysical ghost-like
states \cite{Avdeev1994,Damour1992}. Since this theory admits a
renormalizable interaction with fermions \cite{Avdeev1993,Avdeev1994},
it may produce interesting
applications in particle physics and cosmology \cite{Chizhov2017}.
The symmetries taking place in the flat space were recently discussed
in \cite{Thierry-Mieg2023}. In curved spacetime, the renormalizability
is more restrictive, but still holds owing to the local conformal
symmetry \cite{Cheshire}.

The antisymmetric field model \cite{Avdeev1993,Cheshire} has the
Hamiltonian unbounded from below, indicating possible instabilities
and violation of unitarity. The multiplicative
renormalizability of this model depends on the conformal symmetry.
On the other hand, this symmetry is known to be anomalous at the
quantum level. In what follows, we recalculate the fermionic
contributions in curved space \cite{Cheshire} using two different
ways of doubling for the Dirac operator and compare the results
for both anomaly and the anomaly-induced effective action. The
difference in the local parts of these actions is owing to the local
multiplicative
anomaly, qualitatively similar to the one recently discussed in
\cite{AtA}.

The paper is organized as follows. In Sec.~\ref{sec2}, we review the
fermionic conformal model with an antisymmetric field, following
our previous work \cite{Cheshire}. Sec.~\ref{sec3} reports on the
two schemes of doubling of the curved-space Dirac operator with
the background antisymmetric field and the corresponding one-loop
divergences. Sec.~\ref{sec4} is devoted to the conformal anomaly,
induced action of external fields, consequent ambiguities, and MA.
In the last Sect.~\ref{Conc} we draw our conclusions and discuss
possible extensions of this work.

The conventions include the signature $(+,-,-,-)$, but Wick
rotation to the Euclidean space is assumed in the heat-kernel
calculations. The definition of  the Riemann tensor is $\,R_{\,.\,\be\mu\nu}^{\al}=\Ga_{\,\be\nu,\,\mu}^{\al}
-\Ga_{\,\be\mu,\,\nu}^\al \,+\, ...$,
of the Ricci tensor
$R_{\al\be}=R_{\,.\,\al\mu\be}^\mu$, and the scalar
curvature $R=R_{\,\al}^\al$. Our notations for derivatives are
$\na A = A\na + (\na A)$.

%%%%%%%%%%%%%%%%%%%%%%%%%%%%%%
%%%%%%%%%%%%%%%%%%%%%%%%%%%%%%
%%%%%%%%%%%%%%%%%%%%%%%%%%%%%%
\section{Antisymmetric tensor field with conformal symmetry}
 \label{sec2}

In this work, we do not intend to quantize the antisymmetric tensor
field $B_{\mu\nu}=-B_{\nu\mu}$, but only the Dirac field on the
background of the metric and $B_{\mu\nu}$.\footnote{As we have
pointed out in \cite{Cheshire}, this creates a close analogy with the
semiclassical gravity, where the vacuum action has fourth
derivatives, but this does not imply an inconsistency of the quantum
theory.} At the same time, the corresponding anomaly comes from
the renormalization of the conformal vacuum action, which has to
be properly formulated.

The action of a curved-space theory of the antisymmetric tensor field
$B_{\mu\nu}$, possessing local conformal symmetry in the limit
$m \to 0$, has the form \cite{Cheshire}
\beq
&&
S_B \,=\,
S_g \,+\,
\int d^4x \sqrt{-g}\,\,\Big\{
\frac12\,\big(W_4 + \la W_1\big)
- \frac12M^2 B_{\mu\nu}
- \frac{1}{4!}\,\big(f_2 W_2  + f_3 W_3
\big)
\nn
\\
&&
\qquad \qquad \qquad
\,+\,\,\, \mbox{total derivatives} \Big\}.
\label{actionB}
\eeq
The first term $S_g$ is the metric-dependent vacuum action
(see, e.g., \cite{birdav} and \cite{OUP}),
\beq
S_{HD} &=& \int d^4x \sqrt{-g}\,
\left\{a_1C^2 + a_2E_4 + a_3{\Box}R \right\},
\label{HD}
\eeq
Here $\,C^2=R_{\mu\nu\al\be}^2 - 2 R_{\al\be}^2 + (1/3)\,R^2\,$
is the Weyl tensor square and
$\,E_4 = R_{\mu\nu\al\be}^2 - 4 R_{\al\be}^2 + R^2\,$
is the integrand of the Gauss-Bonnet topological term.
In the $B_{\mu\nu}$-dependent sector, $\la$ is a nonminimal parameter
of the interaction with the Weyl tensor and $f_{2,3}$ are quartic
self-couplings of the antisymmetric field.

The irreducible conformal terms which are building blocks of
the action (\ref{actionB}) are
\beq
&&
W_1 = \sqrt{-g}  \,B^{\mu\nu}B^{\al\be}C_{\al\be\mu\nu},
\nn
\\
&&
W_2 = \sqrt{-g} \,(B_{\mu\nu}B^{\mu\nu})^2,
\nn
\\
&&
W_3 = \sqrt{-g} \, B_{\mu\nu}B^{\nu\al}B_{\al\be}B^{\be\mu},
\nn
\\
&&
W_4 \,=\,
\sqrt{-g}\Big\{
(\na_\al B_{\mu\nu}) (\na^\al B^{\mu\nu})
- 4 (\na_\mu B^{\mu\nu}) (\na^\al B_{\al\nu})
\nn
\\
&&
\qquad
\quad
+ \,2 B^{\mu\nu}R_\nu^{\,\al}B_{\mu\al}
%% + \,2 B^{\mu\nu}B^{\al\be}R_{\mu\al}g_{\nu\be}
-\frac16 \,RB_{\mu\nu}B^{\mu\nu}
\Big\}.
\label{W1234}
\eeq
The reduction formulas for other conformal and nonconformal
terms are listed in Appendix A and the conformal transformations
of these terms in Appendix B.

The rules of conformal transformation for the
metric and for the $B_{\mu\nu}$ field are
\beq
g_{\mu\nu} = \bar{g}_{\mu\nu}\,e^{2\si}\,,
\qquad
B_{\mu\nu} = \bar{B}_{\mu\nu}\,e^{\si}\,,
\qquad
\si = \si(x)\,.
\label{confBg}
\eeq
Since the indices are raised and lowered using the metric,
$B^{\mu\nu} = \bar{B}^{\mu\nu}\,e^{-3\si}$.

%%%%%%%%%%%%%%%%%
The total derivative terms in the action (\ref{actionB}) are
essential for our consideration, different from the previous paper
\cite{Cheshire}. Those include three relevant $B_{\mu\nu}$-dependent
terms
\beq
N_1\,=\,\cx \big(B_{\mu\nu}\big)^2\,,
\quad
N_2 \,=\,\na_\mu \big[B^{\mu\nu} \big(\na^\al B_{\al\nu}\big)\big]
\quad
\mbox{and}
\quad
N_3 \,=\,\na_\mu \big[B_{\al\nu}\big(\na^\al B^{\mu\nu} \big)\big]\,.
\label{Ns}
\eeq

One can conformally couple $B_{\mu\nu}$ to the Dirac fermion
\cite{Avdeev1993,Cheshire} in the form
\beq
S_{1/2}
\,\,=\,\, i\int d^4x\,\sqrt{-g}\,\,\bar{\psi}
\big\{\ga^\mu\na_\mu  \,-\, \Sigma^{\mu\nu}B_{\mu\nu} - im
\big\} \psi,
\label{action-Dirac-B}
\eeq
where $\gamma$-matrices are defined as
$\ga^\mu = e^\mu_{\,a}\ga^a$, \ $\Si^{\mu\nu}
= \frac{i}{2}(\ga^\mu \ga^\nu - \ga^\nu \ga^\mu)$, $m$ is the
mass of the spinor field. A non-zero mass violates
conformal symmetry, but we include it for generality since the
massless limit is smooth.
On the other hand, the massless version of the theory
possesses conformal symmetry under (\ref{confBg}) and the
standard transformations for the fermions,
\beq
\psi = \psi_*\,e^{- \frac32 \,\si}\,,
\qquad
\bar{\psi} = \bar{\psi}_*\,e^{- \frac32 \,\si}\,.
\label{conf_ferm}
\eeq

According to \cite{tmf}, the conformal symmetry
holds in the one-loop counterterms. Therefore, in the massless
case, the one-loop divergences should be of the form (\ref{W1234})
plus surface terms. In the presence of the mass term, the violation
of the local conformal symmetry is soft \cite{Boulware_70}, that has
the same effect in the curved  spacetime \cite{DerVecScal}. In our
present case,  the mass-independent one-loop divergences
has to be  those of the massless theory, i.e., linear
combinations of the terms (\ref{W1234}) and (\ref{HD}), and
the surface terms, such as the integrals of  (\ref{Ns}).

All these expectations were confirmed by the direct one-loop
calculation \cite{Cheshire}, but a few relevant questions remain
open. One of them is about the relationship between the conformal
invariant and the gauge-invariant nonconformal model of
\cite{OgiPolu67}. This part is beyond the scope of the present
work. In the present paper, we explore the ambiguities in the
one-loop divergences of the massless conformal version of the
theory (\ref{action-Dirac-B}) and the corresponding uncertainty
in the trace anomaly and in the finite part of the effective action
of the theory (\ref{actionB}) which results from the path integral
over the fermions.

%%%%%%%%%%%%%%%%%%%%%%%%%%%%%%
%%%%%%%%%%%%%%%%%%%%%%%%%%%%%%
%%%%%%%%%%%%%%%%%%%%%%%%%%%%%%
\section{One-loop divergences for the fermion field}
 \label{sec3}

The purpose of this section is to derive the one-loop divergences
for the Dirac fermion (\ref{action-Dirac-B}) on the background of
external metric and antisymmetric field ${B}_{\mu\nu}$. To evaluate
the divergent part of the functional determinant
\beq
&&
\bar{\Ga}(g,B)
\,=\, - \,i \Tr \log \hat{H},
\label{barGa}
\\
&&
\hat{H}  \,=\,
\ga^\mu\na_\mu - \Si_{\mu\nu} B_{\mu\nu} + im\,,
\label{H}
\eeq
we need a doubling procedure reducing the operator to the standard
form. For this, we need a conjugate operator $ \hat{H}^*$. Let us
consider the following two choices:
\beq
&&
\hat{F}_1 \,=\, \hat{H} \hat{H}_1^*,
\qquad
\hat{H}_1^*
\,=\,
 \ga^\mu\na_\mu - \Si^{\mu\nu} B_{\mu\nu} - im,
\label{Hstar1}
\\
&&
\hat{F}_2 \,=\, \hat{H} \hat{H}_2^*,
\qquad
\hat{H}_2^*
\,=\,
\ga^\mu\na_\mu - im.
\label{Hstar2}
\eeq
The first choice was elaborated in \cite{Cheshire}. Since
$\Tr \log \hat{H} = \Tr \log \hat{H}_1^*$, we can use the relation
\beq
- \,i \Tr \log \hat{H}\,=\, - \,\frac{i}{2}\, \Tr \log \hat{F}_1.
\label{GaF1}
\eeq
For the second choice, we note that $\hat{H}_2^*$ does not
depend on the field ${B}_{\mu\nu}$. Therefore, for the
${B}_{\mu\nu}$-independent part of effective action we can use the
same relation (\ref{GaF1}). Indeed, this part is pretty well-known
(see, e.g., \cite{OUP}) and we can skip it and concentrate on the
${B}_{\mu\nu}$-independent part, which obeys the rule
\beq
- \,i \Tr \log \hat{H}\,=\, - \,i\, \Tr \log \hat{F}_2.
\label{GaF2}
\eeq
Both operators have the standard form
\beq
\hat{F}_k \,=\, \hat{H} \hat{H}_k^*
\,=\,
\hat{1}\cx + 2\hat{h}_k^\al \na_\al + \hat{\Pi}_k ,
\qquad
k = 1,2.
\label{Fk}
\eeq
The elements of the two operators are
\beq
&&
\hat{h}_1^\al \,=\, 2 \ga^5 \ga_\be \tilde{B}^{\al\be},
\label{hPi1}
\\
&&
\hat{\Pi}_1\,=\, m^2 - \frac14\,R + 2B_{\al\be} B^{\al\be}
- 2i (\na_\al B^{\al\be})\ga_\be
- 2i \ga^5 B_{\al\be} \tilde{B}^{\al\be}
+ 2 \ga^5  (\na_\al \tilde{B}^{\al\be})\ga_\be
\nn
\\
&&
\mbox{and}
\quad
\hat{h}_2^\al \,=\, i \ga_\be B^{\al\be}
\,+\,
\ga^5 \ga_\be \tilde{B}^{\al\be},
\nn
\\
&&
\qquad \,\,\,
\hat{\Pi}_2\,=\, m^2 - \frac14\,R + im B_{\al\be} \Si^{\al\be},
\label{hPi2}
\eeq
where the dual tensor is defined as
\beq
\tilde{B}_{\mu\nu}
\,=\,\frac12\,\vp_{\mu\nu\al\be} B^{\al\be}\,.
\label{dualB}
\eeq

The one-loop divergences can be derived using the standard
heat-kernel technique \cite{DeWitt65}. For the first scheme
(\ref{Hstar1}) one can find full details in \cite{Cheshire} and the
calculation in the second scheme case is technically similar. For
the sake of generality, we present the results for the massive field,
however later on set $m=0$. The full set of reduction formulas can
be found in Appendix A, so let us directly give the formulas for
one-loop divergences,
\beq
&&
{\bar \Ga}^{(1)}_{div,\,k}
\,=\,{\bar \Ga}^{(1)}_{div}(g)\,+\,{\bar \Ga}^{(1)}_{div,\,k}(B),
\qquad
k=1,2;
\label{EA-g-divs}
\\
&&
{\bar \Ga}^{(1)}_{div}(g)\,=\,
-\,\frac{\mu^{n-4}}{\vp}\int d^nx\sqrt{-g}\,
\Big\{\frac{1}{20}\,C_{\mu\nu\al\be}^2
- \frac{11}{360}\,E_4
+ \frac{1}{30}\,{\Box}R
+ \frac13\,m^2 R
- 2m^4
\Big\},
\nn
\\
&&
{\bar \Ga}^{(1)}_{div,\,1}(B)\,=\,
-\,\frac{\mu^{n-4}}{\vp}\int d^nx\sqrt{-g}\,
\Big\{\frac43\,\big(W_1 - W_4 - 2 W_2 + 8 W_3 \big)
+ \frac83\,N_1 %%  {\cx} B^2_{\mu\nu}
+ 8m^2 B^2_{\mu\nu}\Big\},
\nn
\\
&&
\qquad \qquad \qquad
\label{EA-B1}
\\
&&
{\bar \Ga}^{(1)}_{div,\,2}(B)\,=\,
-\,\frac{\mu^{n-4}}{\vp}\int d^nx\sqrt{-g}\,
\Big\{\frac43\,\big(W_1 - W_4 - 2 W_2 + 8 W_3 \big)
\nn
\\
&&
\qquad \qquad \qquad
+ \,\,\frac43\,\big(N_3 - N_2\big)
+ 8m^2 B^2_{\mu\nu}\Big\},
\label{EA-B2}
\eeq
where $\vp = (4\pi)^2(n-4)$ is the parameter of dimensional
regularization.

The two expressions (\ref{EA-B1}) and (\ref{EA-B2}) demonstrate
the conformal invariance of the coefficient of the $1/\vp$ pole
in the limit $m\to 0$ and $n\to 4$. Since this follows from the
general theorem proved in \cite{tmf} (see also \cite{OUP}
for the introductory-level consideration of the simplest case which is
sufficient here), which means that these formulas passed the basic test
of correctness. On the other hand, according to the relations
(\ref{GaF1}) and (\ref{GaF2}), the two expressions
${\bar \Ga}^{(1)}_{div,\,1}(B)$ and
${\bar \Ga}^{(1)}_{div,\,2}(B)$ should be equal, but this is not
exactly true. It is easy to note that Eqs.~(\ref{EA-B1}) and
(\ref{EA-B2}) differ by the total derivative terms. These terms do
not have relevance by their own, but their difference produces an
ambiguity in the anomaly, which we discuss in the next section.

%%%%%%%%%%%%%%%%%%%%%%%%%%%%%%
%%%%%%%%%%%%%%%%%%%%%%%%%%%%%%
%%%%%%%%%%%%%%%%%%%%%%%%%%%%%%
\section{Anomaly and anomaly-induced action}
 \label{sec4}

The trace anomaly is the violation of
Noether identity corresponding to the local conformal symmetry.
At the classical level, this identity has the form corresponding to
(\ref{confBg}) and (\ref{conf_ferm}),
\beq
\frac32\, \bigg(
\bar{\psi}\,\frac{\de S_c}{\de \bar{\psi}}
\, + \, \frac{\de S_c}{\de \psi}\,\psi\bigg)
-\,2\, g_{\mu\nu}\,\frac{\de S_c}{\de g_{\mu\nu}}
\,-\,B_{\mu\nu}\,\frac{\de S_c}{\de B_{\mu\nu}}
\, = \, 0 \,,
\label{Noether-conf}
\eeq
where the conformal action is \ $S_c = S_B + S_{1/2}$ \ with $M=m=0$
in the expressions (\ref{actionB}) and (\ref{action-Dirac-B}) for the
actions of background fields $B_{\mu\nu}$ and  $g_{\mu\nu}$;
$S_{1/2}$  is the action of the quantum fields $\bar{\psi}$ and $\psi$.
It would be interesting to extend the consideration to the quantum field
$B_{\mu\nu}$ and to explore its contribution to the anomaly, especially
in the purely gravitational sector. However, there is a technical
obstacle, i.e.,  the unknown contribution to divergences from the
nonminimal operator in the space of antisymmetric fields. So, in
\cite{Cheshire} and in the present work, we restrict the consideration
by the quantum effects of fermions. Then the anomaly emerges only in
the vacuum part of the effective action and we need to evaluate
\beq
\langle
\mathcal{T}
\rangle
\,\,=\,\,
-\,\frac{2}{\sqrt{-g}}\, g_{\mu\nu}\,\frac{\de \Ga(g,B)}{\de g_{\mu\nu}}
\,-\,\frac{1}{\sqrt{-g}}\,B_{\mu\nu}\,\frac{\de \Ga(g,B)}{\de B_{\mu\nu}}.
\label{anomaly}
\eeq
In this equation, $ \Ga(g,B)$ is the renormalized effective action
of the fields $B_{\mu\nu}$ and  $g_{\mu\nu}$. At the one-loop level,
this effective action is a sum of the classical action, divergent and
finite parts of the loop contribution, and the divergent local
counterterm required to make the sum finite. The anomaly comes from
the finite part of the loop corrections and does not depend on the
regularization (see the discussion in \cite{OUP}). However, the
easiest way to arrive at the anomaly is by using the
dimensional regularization and the locality of the counterterms
\cite{duff77}. In this way, there are no ambiguities in the nonlocal
part of the anomaly-induced action (which is the aforementioned
finite part) because this part is nothing but the mapping from the
leading logarithmic contribution to the polarization operator, or to
the effective potential of the background fields. However, there may
be an ambiguity in the local part because it is not related
to the leading logarithms and relevant divergences.

Let us now see how it works in our case.
The anomaly derived in the standard way \cite{duff77,OUP}
repeats the form of divergences (\ref{EA-g-divs}) with $m=0$,
\beq
&&
\langle
\mathcal{T}
\rangle
\,\,=\,\,-\,
\frac{1}{(4\pi)^2}\,\bigg\{
\frac{1}{20}\,C_{\mu\nu\al\be}^2
- \frac{11}{360}\,E_4
+ \frac{1}{30}\,{\Box}R
\nn
\\
&&
\qquad
+ \,\,
\frac43\,\big[
W_1 - W_4 - 2 W_2 + 8 W_3\big]
\,+\, \ga_1N_1  + \ga_2 (N_3-N_2)
\bigg\}.
\label{T}
\eeq
In this expression
\beq
&&
\ga_1\,=\,\frac43\,,\qquad \ga_2\,=\,0
\quad
\mbox{for the scheme of doubling (\ref{Hstar1})};
\label{ga1}
\\
&&
\ga_1\,=\,0\,,
\qquad \,
\ga_2\,=\,\frac83
\quad
\mbox{for the scheme of doubling (\ref{Hstar2})}.
\label{ga2}
\eeq
It is worth noting that we assumed the value of the coefficient of
the $\cx R$-term corresponding to all regularizations except the
dimensional one and the covariant Pauli-Villars, where this beta
function is ambiguous \cite{anomaly-2004}. Different from this
case, the divergence between (\ref{ga1}) and  (\ref{ga2}) is not
related to the choice of regularization.

To clarify the difference between the two expressions for the
anomalies, consider the anomaly-induced action.
This action consists of the nonlocal and local terms. The treatment
of nonlocal ones is pretty much standard (see, e.g., \cite{OUP,AtA}),
but we briefly describe it here for completeness. First of all, let us
introduce a collective notation for the legitimate conformal terms
($C$-invariants) in Eq.~(\ref{T}),
\beq
Y \,\,=\,\,
\frac{1}{(4\pi)^2}\,\bigg\{
\frac{1}{20}\,C_{\mu\nu\al\be}^2
\,+ \,
\frac43\,\big[
W_1 - W_4 - 2 W_2 + 8 W_3\big]\bigg\}.
\label{Y}
\eeq
The next step is to remember the conformal rule for the
modified topological term \cite{rie,FrTs84}
\beq
&&
\sqrt{-g}\,\Big(E_4-\frac23\,{\cx} R\Big)
\,=\,
\sqrt{-\bar{g}}\,\Big({\bar E_4}-\frac23\,{\bar \cx} {\bar R}
+ 4{\bar \De_4}\si \Big),
\label{119}
\\
&&
\mbox{where}
\quad
\Delta_4 \,=\,
\cx^2 + 2R^{\mu\nu}\nabla_{\mu}\nabla_{\nu} - \dfrac{2}{3}R\cx
+\dfrac{1}{3}(\nabla^{\mu}R)\nabla_{\mu},
\eeq
with
$\sqrt{-g}\Delta_4=\sqrt{-\bar{g}}\bar{\Delta}_4$
\cite{FrTs-superconf,Paneitz}. On top of this we need a special
notation for the coefficient of the Gauss-Bonnet term in (\ref{T}),
\beq
&&
b\,=\,-\,\frac{11}{360\,(4\pi)^2}.
\label{b}
\eeq
After this, the non-local term can be written in a universal form
(see, e.g., \cite{OUP,AtA})
\beq
&&
\Ga_{ind,\,nonloc}\,\,=\,\,
\frac{b}{8}\int_x\int_y\Big(E_4
-\frac23\square R\Big)_{\hspace{-1mm}x}
G(x,y)\Big(E_4-\frac23\square R\Big)_{\hspace{-1mm}y}
\nn
\\
&&
\qquad \qquad \qquad
+\,\,
\frac{1}{4}\int_x\int_y Y(x)\, G(x,y)
\Big(E_4-\frac23\square R\Big)_{\hspace{-1mm}y},
\label{nonlocal-S}
\eeq
where we used $\int_x\equiv \int d^4 x\sqrt{-g(x)}$
and the Green function of the Paneitz operator
\beq
(\sqrt{-g}\De_4)_xG(x,y)\,=\,\de(x,y).
\label{Green_function}
\eeq
Now we consider the integration of the remaining total derivative
terms in the anomaly. There is a general belief that for each such
term there is a local term in the anomaly-induced action, regardless
(up to our knowledge) there is no proof that this is always the case.
For the $\cx R$-term the solution is known from the formula
\beq
&&
-\frac{2}{\sqrt{-g}}g_{\mu\nu}\frac{\delta}{\delta g_{\mu\nu}}
\int_x R^2\,=\,12\square R.
\label{TrR2}
\eeq
Thus, it remains to integrate the total derivative terms with
$\ga_1$ and $\ga_2$ in (\ref{T}). Using the results for the conformal
transformations collected in Appendix B, we obtain
\beq
&&
-\frac{2}{\sqrt{-g}}g_{\mu\nu}\frac{\delta}{\delta g_{\mu\nu}}
\int_x R B^2_{\mu\nu}\,=\,6 N_1 \,,   %% \cx B^2_{\mu\nu}
\label{Trga1}
\\
&&
-\frac{2}{\sqrt{-g}}g_{\mu\nu}\frac{\delta}{\delta g_{\mu\nu}}
\int_x \big( \na_\al B_{\mu\nu}\big)^2 \,=\,N_3 - N_2\,.
\label{Trga2}
\eeq

Thus, the local terms in the anomaly-induced actions are as follows:
\beq
&&
\Ga^{(1)}_{ind,\,\ga 1}
\,\,=\,\,
-\,\, \frac{\ga_1}{6(4\pi)^2}\int_x %% d^4x\sqrt{-g}
\,R B_{\mu\nu}B^{\mu\nu}\,,
\label{Gaga1}
\\
&&
\Ga^{(1)}_{ind,\,\ga 2}
\,\,=\,\,
\frac{\ga_2}{12(4\pi)^2} \int_x %%  d^4x \sqrt{-g}
\,\big\{3\, \big( \na_\al B_{\mu\nu}\big)^2
\,-\,2\,R B_{\mu\nu}B^{\mu\nu}\big\}\,.
\label{Gaga2}
\eeq
It is important to stress that both these expressions are subjects of
an extra ambiguity because we can add to the integrands the conformal
invariant  $W_4$ from (\ref{W1234}) with arbitrary coefficients.
However, since this invariant is not a combination of the integrands
of (\ref{Gaga1}) and  (\ref{Gaga2}), this operation cannot eliminate
the difference between the two local functionals.

The full expression of the anomaly-induced effective action is
\beq
\Ga_{ind}\,\,=\,\,S_c(g,B) \,+\,\Ga_{ind,\,nonloc}
\,+\, \Ga^{(1)}_{ind,\,\ga 1} \,+\, \Ga^{(1)}_{ind,\,\ga 2}
\,+\,\frac{7}{540(4\pi)^2}\int_x R^2\,.
\label{Gaind}
\eeq
where $Sc(g,B)$ is an arbitrary conformally invariant functional of
the fields $g_{\mu\nu}$ and  $B_{\mu\nu}$, which is an integration
constant to Eq.~(\ref{anomaly}). As we just noted, the local terms
may be changed by adding the $W_4$ term to $Sc(g,B)$. However,
even after doing this, the difference between the terms
$\Ga^{(1)}_{ind,\,\ga 1}$ and $\Ga^{(1)}_{ind,\,\ga 2}$ does not
vanish, indicating the presence of a local multiplicative anomaly
that does not depend on the renormalization conditions.

%%%%%%%%%%%%%%%%%%%%%%%%%%%
%%%%%%%%%%%%%%%%%%%%%%%%%%%

%%%%%%%%%%%%%%%%%%%%%%%%%%%
%%%%%%%%%%%%%%%%%%%%%%%%%%%
%%%%%%%%%%%%%%%%%%%%%%%%%%%
\section{Conclusions and discussions}
\label{Conc}

The trace anomaly is well-known to have an ambiguity related
to the total derivative $\cx R$ term, which results in the ambiguity
of the local $R^2$ term in the anomaly-induced effective action.
The origin of this ambiguity, as well as the similar one with
$\cx \ph^2$-term in the theories with external scalar field $\ph$,
can be attributed to the peculiarities in the choice of regularization
scheme, such as dimensional \cite{birdav,duff77}, or the covariant
Pauli-Villars  \cite{anomaly-2004,anomaly-2006} regularizations.
Is it true that the ambiguity in the total derivative terms in the
anomaly may be related only to the choice of regularization? In the
recent paper \cite{AtA}, we found an example of the opposite.
For the quantum fermion field, when the background fields include
metric and torsion, the ambiguity comes from the different schemes
of doubling of the Dirac operator and does not depend on the
choice of regularization. Here we present one more example of the
same sort, this time with the background metric and antisymmetric
field. The calculations in this case are more complicated and our
present work serves also as verification of the previous result in
\cite{Cheshire}, where the derivation of one-loop divergences is an
important ingredient of the general analysis of renormalizability of
the Avdeev and Chizhov model \cite{Avdeev1993} in curved spacetime.

 The calculation of divergences in the two different schemes of
 fermion doubling confirmed the main conclusion of \cite{Cheshire}
 concerning the renormalizability of the conformal theory
 \cite{Branson} coupled to fermions, including when this symmetry
 is softly broken by the masses. On the other hand, there is a
 difference in the two ways of calculation, which concerns the total
 derivative terms. These terms do not violate conformal invariance,
 that is the Noether identity (\ref{Noether-conf}) for the coefficient
 of the pole in the one-loop divergences. On the other hand, there are
 ambiguities in the local non-conformal terms in the anomaly-induced
 effective action caused by two different total derivative terms in
 the divergences and anomalies. This ambiguity is also the second
example of the local multiplicative anomaly, similar to the one
discussed in the case of torsion \cite{AtA}.

It would be interesting to extend our analysis to the complete
interacting theory, that is quantize not only the fermions but also
the antisymmetric field. The main obstacle in this way is the
proper-time representation of the propagator of the nonminimal
field without gauge symmetry, which follows from the $W_4$
term in (\ref{W1234}). We hope to have progress in solving this
problem in the near future.

%%%%%%%%%%%%%%%%%%%%%%%%%%%%%%%
%%%%%%%%%%%%%%%%%%%%%%%%%%%%%%%
%%%%%%%%%%%%%%%%%%%%%%%%%%%%%%%
\section*{Acknowledgements}

The author is grateful to I.L. Buchbinder for the discussion of
the Hamiltonian and possible instabilities in the conformal
antisymmetric field model. The partial support from CNPq
(Conselho Nacional de Desenvolvimento Cient\'{i}fico e
Tecnol\'{o}gico, Brazil) under the grant 305122/2023-1, in
gratefully acknowledged.

%%%%%%%%%%%%%%%%%%%%%%%%%%%
%%%%%%%%%%%%%%%%%%%%%%%%%%%
%% \appendixes
%%%%%%%%%%%%%%%%%%%%%%%%%%%

%%%%%%%%%%%%%%%%%%%%%%%%%%%
\section*{Appendix A. Basic reduction formulas}

The formulas listed below are more general than the ones in
Ref.~\cite{Cheshire} because they include total derivative terms.

The initial relation is a version of the first formula of (\ref{W1234}),
\beq
W_{11} = \sqrt{-g} \, B^{\mu\al}B^{\nu\be}C_{\al\be\mu\nu}
= \frac12\,W_1.
\label{W11red}
\eeq
Other basic definitions include (\ref{Ns}) and
\beq
&&
K_1 = \sqrt{-g} \,B^{\mu\nu}B^{\al\be} R_{\mu\al} g_{\nu\be},
\qquad
K_2 = \sqrt{-g} \, B_{\mu\nu}B^{\mu\nu}R,
\nn
\\
&&
K_3 = \sqrt{-g} \,(\na_\al B_{\mu\nu}) (\na^\al B^{\mu\nu})
 = \sqrt{-g} \,(\na_\al B_{\mu\nu})^2,
\nn
\\
&&
K_4 = \sqrt{-g} \,(\na_\mu B^{\mu\nu}) (\na^\al B_{\al\nu})
= \sqrt{-g} \,(\na_\mu B^{\mu\nu})^2 .
\label{K123}
\eeq
The reduction formulas are
\beq
&&
K_{11} = \sqrt{-g} \, B^{\mu\nu}B^{\al\be}R_{\mu\nu\al\be}
= 2K_1 - \frac13\,K_2 + W_1.
\nn
\\
&&
K_{12} = \sqrt{-g} \, B^{\mu\al}B^{\nu\be}R_{\mu\nu\al\be}
= \frac12\,K_{11}.
\nn
\\
&&
%%  Z_4
K_{31} = \sqrt{-g} \,(\na_\al B_{\mu\nu}) (\na^\mu B^{\al\nu})
= K_4 - \frac16\,K_2 + \frac12\,W_1 + N_2 - N_3 \,.
\label{K31red}
\eeq
%%%%%%%%%%%%%%%%%   From old
%%%%%%%%%%%%%%%%%   From old
%%%%%%%%%%%%%%%%%   From old
%%%%%%%%%%%%%%%%%   From old

The next set of formulas involves $\tilde{B}_{\mu\nu}$. Those
are derived using the contractions of two antisymmetric tensors,
e.g., $\vp^{\mu\nu\al\be}\vp_{\mu\nu\rho\si}
= -2 \big(\de^\al_{\,\rho}\de^\be_{\,\si}
- \de^\al_{\,\si}\de^\be_{\,\rho}\big)$.
The initial relation is
\beq
&&
\tilde{B}_{\mu\nu}\tilde{B}^{\al\be}
\,=\,
- \,B_{\mu\nu}B^{\al\be}
\,-\,
\frac12\,B_{\rho\si}^2
\big(\de_\mu^\al \, \de_\nu^\be
- \de_\nu^\al \, \de_\mu^\be\big)
\nn
\\
&&
\qquad \qquad\,\,
+\,\,\,
\de_\mu^\al B_{\nu\la} B^{\be\la}
- \de_\nu^\al B_{\mu\la} B^{\be\la}
+ \de_\nu^\be B_{\mu\la} B^{\al\la}
- \de_\mu^\be B_{\nu\la} B^{\al\la}\,.
\label{til1}
\eeq
The contractions are
\beq
&&
\tilde{B}^{\mu\nu}\tilde{B}^{\al\be} g_{\nu\be}
\,=\,
B^{\mu\nu}B^{\al\be} g_{\nu\be}
- \frac12\, B^{\rho\si}B_{\rho\si} g^{\mu\al},
\nn
\\
&&
\tilde{B}^{\mu\nu}\tilde{B}_{\mu\nu}
= - B^{\mu\nu}B_{\mu\nu},
\label{til2}
\eeq
which also gives
\beq
&&
C_{\al\be\mu\nu} \tilde{B}^{\al\be}\tilde{B}^{\mu\nu}
\,=\,
-\, C_{\al\be\mu\nu} B^{\al\be} B^{\mu\nu}\,=\, - \,W_1 \,,
\nn
\\
&&
R_{\al\be\mu\nu} \tilde{B}^{\al\be}\tilde{B}^{\mu\nu}
\,=\,
2 R_{\al\be\mu\nu} \tilde{B}^{\al\mu}\tilde{B}^{\be\nu}
\,=\,
- \,W_1 + 2K_1 - \frac23\,K_2\, .
\label{til4}
\eeq

Further relations include
\beq
&&
(\na_\al \tilde{B}_{\mu\nu}) (\na^\al \tilde{B}^{\mu\nu})
\,=\, - \,K_3,
\nn
\\
&&
(\na_\mu \tilde{B}^{\mu\nu}) (\na^\al \tilde{B}_{\al\nu})
\,=\,
K_4  - \frac12\, K_3 - \frac16\,K_2 + \frac12\, W_1 + N_2 - N_3\,,
\nn
\\
&&
(\na_\al \tilde{B}_{\mu\nu}) (\na^\mu \tilde{B}^ {\al\nu})
\,=\,
- \,\frac12\, K_3 + K_4
\label{til5}
\eeq
and
\beq
&&
(\tilde{B}_{\mu\nu} \tilde{B}^{\mu\nu})^2
\,=\, W_2\,,
\nn
\\
&&
\tilde{B}_{\mu\nu} B^{\mu\nu}
\tilde{B}_{\al\be} B^{\al\be}
\,=\,
-\,2\,W_2  \,+\,4\,W_3\,,
\nn
\\
&&
\tilde{B}_{\mu\nu} \tilde{B}^{\nu\al}
\tilde{B}_{\al\be} \tilde{B}^{\be\mu}
\,=\, W_3\,.
\label{til7}
\eeq

%%%%%%%%%%%%%%%%%%%%%%%%%%%%%%%%%%
\section*{Appendix B. Conformal variations of local terms}

Let us first list the infinitesimal conformal variations of the
irreducible terms \cite{Cheshire} used in the main text. The
basic variations (see, e.g., \cite{Stud}) are
\beq
&&
\de_c  \Ga^\la_{\al\be}
\,=\, \de^\la_{\,\al}\si_\be + \de^\la_{\,\be}\si_\al
- \bar{g}_{\al\be} \si^\la,
\nn
\\
&&
\de_c  R \,=\,  - 2\bar{R}\si - 6 \bar{\cx}\si,
\nn
\\
&&
\de_c  R_{\al\be}
\,=\,  - \bar{g}_{\al\be} \bar{\cx}\si - 2 \si_{\al\be},
\label{conbasic}
\eeq
where $\si_\al = \bar{\na}_\al \si$,  \
$\si^\al = \bar{g}^{\al\be}\si_\be$, \ and \
$\si_{\al\be} = \bar{\na}_\al \bar{\na}_\be \,\si$.
The covariant derivatives with bars correspond to the
fiducial metric $\bar{g}_{\al\be}$.
The variations of the terms (\ref{K123}) are
\beq
&&
\de_c K_1 = \sqrt{- \bar{g}}
\, \bar{B}^{\mu\nu}
\big[
   2 \si^\la (\bar{\na}_\la \bar{B}_{\mu\nu})
+ 2 \si_\nu (\bar{\na}^\la \bar{B}_{\mu\la})
+ 2 \si^\la (\bar{\na}_\nu \bar{B}_{\mu\la})
\big],
\nn
\\
&&
\de_c K_2 = \sqrt{-\bar{g}} \, \bar{B}^{\mu\nu}
\big[12  \si^\la (\bar{\na}_\la \bar{B}_{\mu\nu})\big],
\nn
\\
&&
\de_c K_3 = \sqrt{-\bar{g}} \,\bar{B}^{\mu\nu}
\big[
4 \si_\nu (\bar{\na}^\la \bar{B}_{\mu\la})
- 4 \si^\la (\bar{\na}_\nu \bar{B}_{\mu\la})
- 2 \si^\la (\bar{\na}_\la \bar{B}_{\mu\nu})
\big],
\nn
\\
&&
\de_c K_4 = \sqrt{-\bar{g}} \,\bar{B}^{\mu\nu}
\,\big[2 \si_\nu (\bar{\na}^\la \bar{B}_{\mu\la})\big].
\label{deltasK123}
\eeq
Using these relations and some additional algebra, we get
the expression of variations in terms of the surface terms,
including (\ref{Ns}). For instance, one can easily get
\beq
&&
-\,\frac{2}{\sqrt{-g}}\,g_{\al\be}\,\frac{\de}{\de g_{\al\be}}
\,\int d^4x \sqrt{-g}\,K_2 \,\,=\,\, 6 N_1\,.
\label{confaK2}
\eeq
Another way to arrive at the same formula is to ignore the total
derivative term in the second formula of (\ref{deltasK123}), that
gives an equivalent result
\beq
&&
\de_c K_2 \,\,=\,\, -\,6 \si N_1\,.
\label{confaK2-prime}
\eeq
Similar operations can be applied to other three terms to get
\beq
&&
\de_c K_1 \,\,=\,\,
-\,\si \big(N_1  \,+\, 2 N_2  \,+\, 2N_3\big)\,,
\nn
\\
&&
\de_c K_3 \,\,=\,\, \si \big(N_1  \,-\, 4 N_2  \,+\, 4N_3\big)\,,
\nn
\\
&&
\de_c K_4 \,\,=\,\, -2 \si N_2\,.
\label{confas}
\eeq
These relations can be presented in the form similar to
Eq.~(\ref{confaK2}).

%%%%%%%%%%%%%%%%%%%%%%%%%%%%%%%
%%%%%%%%%%%%%%%%%%%%%%%%%%%%%%%
%%%%%%%%%%%%%%%%%%%%%%%%%%%%%%%

\end{document}